\input{aipcheck}

\documentclass[
    ,final            
  ]
  {aipproc}

\layoutstyle{6x9}

\begin{document}

\title{Precision X-ray measurements on kaonic atoms at LNF}



\author{Johann Marton for the SIDDHARTA Collaboration\footnote{
SIDDHARTA Collaboration at LNF,
http://www.lnf.infn.it/esperimenti/siddharta/ }}{
  address={Stefan Meyer Institut (SMI), Austrian Academy of Sciences,
  Boltzmanngasse 3, 1090 Vienna, Austria}
}



\begin{abstract}

After the successfully performed DEAR experiment at DA$\Phi$NE -
resulting in the most precise data on the hadronic shift and width
in kaonic hydrogen up-to-now - the next step will be the
measurement at the percent level using new X-ray detectors. These
detectors (silicon drift detectors) are developed within the
SIDDHARTA project. The asynchronous background will be suppressed
using the time correlation between the kaon and the X-ray by 2-3
orders of magnitude. These measurements will lead to precise
values of the isospin-dependent antikaon-nucleon scattering
lengths, thus opening a new insight in the low-energy kaon nucleon
interaction.

\end{abstract}

\maketitle


\section{Introduction}
Exotic atoms are extremely valuable test systems for research in
many fields of physics, like the study of fundamental interactions
(e.g. strong interaction in hadronic atoms) and fundamental
symmetries (e.g. CPT tests with antiprotonic helium and
anti-hydrogen).

The experimental studies of simple hadronic atoms as kaonic
hydrogen - in which the electron is substituted by a negatively
charged kaon - provide the straightforward investigation of the
strong antikaon-nucleon interaction at lowest energy. Whereas the
principal interaction in kaonic hydrogen atoms is electromagnetic
the effect of the strong interaction leads to the experimentally
observable shift $\varepsilon_{1s}$ from the electromagnetic value
and to a width $\Gamma_{1s}$ due to the reduced life time of the
ground state. The electromagnetic value of the K$_{\alpha}$
transition can be calculated by solving the Klein-Gordon equation
taking into account various corrections like vacuum polarization
etc. \cite{Indelicato05}. The physical values of the observables
$\varepsilon_{1s}$ and $\Gamma_{1s}$ can be measured to a high
degree of precision by X-ray spectroscopy of the transitions to
the ground state (K transitions) in kaonic hydrogen and deuterium.

In kaonic hydrogen the $\Lambda(1405)$ sub-threshold resonance - a
4 star object in the listings of the Particle Data Group
\cite{PDG} - leads to the repulsive-type kaon-proton interaction
at threshold \cite{Weise95}. The nature of $\Lambda(1405)$ is
still in discussion. According to theory \cite{akaishi}
$\Lambda(1405)$ can be identified as a K$^{-}$p bound state.
Following this theory the existence of deeply bound kaonic
clusters \cite{Akaishi04} is anticipated. The search for these
kaonic nuclear clusters is currently conducted in experimental
studies at GSI and LNF \cite{FOPI, FINUDA} and in the future at
LNF and J-PARC (Japan) \cite{Amadeus-LOI,JPARC}.

The extraction of the isospin-dependent scattering lengths
requires the knowledge of $\varepsilon_{1s}$ and $\Gamma_{1s}$ of
kaonic hydrogen as well as deuterium. No experimental data on the
strong interaction in kaonic deuterium atoms are available up to
now. The investigation of kaonic deuterium will be more
challenging also due to the anticipated low yield of the X-ray
transitions to the ground state.

An ambitious experimental program devoted to kaonic atoms was
started by the DEAR (DA$\Phi$NE Exotic Atom Research)
collaboration \cite{dearcase}. DA$\Phi$NE - the electron-positron
collider at Laboratori Nazionali di Frascati (LNF) turned out to
be the ideal machine for this kind of research. It delivers nearly
mono-energetic K$^{-}$ mesons (E$_{K}$ $\sim$ 16 MeV) from the
$\Phi$ decay with a branching ratio of $\sim$ 50\%. The DA$\Phi$NE
performance in providing low-energy kaons to be stopped in gas
targets is unique and by far superior compared to accelerators
used in former experiments which produced large hadronic
background. Another advantage was ensured by X-ray detectors
(charge-coupled-devices, CCDs) with significantly higher energy
resolution ($\sim$ factor 2 better than  Si(Li) detectors).
However, the drawback of CCDs is the limitation in the background
suppression. Since CCDs have no timing capability only background
leading to pixel clusters (background from gamma rays or charged
particles) can suppressed by analyzing the hit pattern.

The DEAR experiment was successfully performed. The X-ray spectrum
of kaonic nitrogen was measured and even more important DEAR
obtained the most precise values of the strong interaction shift
and width in kaonic hydrogen up to now.

The continuation of the exotic atom program at LNF will employ
large area silicon drift detectors (SDDs) which are providing
timing capability as well as high energy resolution like CCDs.
These detectors are already produced by the SIDDHARTA (Silicon
Drift Detector for Hadronic Atom Research by Timing Application)
Collaboration. A completely new setup with a large array of these
SDDs was designed and is being near completion.

\section{Experiments and Results}

\subsection{Kaonic atoms at DA$\Phi$NE}

Experimental X-ray studies of kaonic nitrogen and finally of
kaonic hydrogen were performed by the DEAR Collaboration. This
line of experiments verfied that DA$\Phi$NE represents the ideal
site for kaonic atom experiments with stopped kaons. Cryogenic gas
targets with a hydrogen gas density (3\% of liquid density) were
employed to provide sufficient kaon stop efficiency and and low
losses due to Stark effect leading to kaon absorption from higher
s states. The X-ray detection was performed by an array of 16 CCDs
(type CCD-55). The energy calibration was done in-beam using the
fluorescence lines from the structure materials of the experiment.
Extremely important for the kaonic hydrogen measurement was the
treatment of the large soft X-ray background. Background data from
measurements without electron-positron collisions and from the
kaonic nitrogen measurements were used.

\subsection{Kaonic hydrogen - Results}

For the first time kaonic nitrogen X-ray spectra of the kaonic
transitions 5-4, 6-5 and 7-6 (i.e. transitions not influenced by
strong interaction) were measured \cite{ishiwatari04}. These data
are important for the kaonic cascade theory and also for
feasibility studies for future
precision experiments on the charged kaon mass.\\

The DEAR experiment verified the repulsive type of the kaon-proton
interaction found by the KEK experiment \cite{KpX} but obtained
smaller values for $\epsilon_{1s}$ and $\Gamma_{1s}$ with higher
precision \cite{Beer}: $\epsilon_{1s} = - 193 \pm 37(stat.) \pm
6(syst.) \mbox{ eV}$ and $\Gamma_{1s} = 249 \pm 111(stat.) \pm
30(syst.) \mbox{ eV}$. These results stimulated  new theoretical
studies on kaonic hydrogen \cite{theory1,theory2,theory3,theory4}.

\section{The SIDDHARTA experiment}

\subsection{X-ray detector development}

In spite of the fact that in DEAR the most precise values of
$\epsilon_{1s}$ and $\Gamma_{1s}$ were extracted the goal of
precision measurements at the percent level calls for a new
experiment.  The precision of the DEAR results was mainly limited
by the high soft X-ray background leading to a signal-to-noise
ratio of $\sim$1:70. The background consists of a synchronous
component related to the produced kaons (hadronic background) and
asynchronous background coming from electromagnetic showers
produced by beam electrons lost due to Touschek effect or residual
gas interaction. Using a new type of X-ray detectors (SDD) the
limitation by the high asynchronous background will be overcome.

The new SDD detectors provide very good energy resolution
(comparable with CCDs) but also timing capability (see table 1).
The charged kaon pair (K$^{+}$ + K$^{-}$) emitted in the $\Phi$
decay can be used as well suited trigger since the kaons are
nearly monoenergetic and emitted back-to-back. The application of
a triple coincidence between the kaon pair and the X-ray will
suppress efficiently the asynchronous background component by
about 2-3 orders of magnitude.

\begin{table}
  \centering
  \caption{Comparison between the X-ray detectors CCDs and SDDs.}\label{CCD-SDD}
  \vspace{0.5 cm}
  \begin{tabular}{|l|l|l|}
    \hline\hline
   X-ray detector & CCD & SDD \\
    \hline
    Type & CCD55-30 & SIDDHARTA \\
    Active area  [mm$^{2}$] & 724  &  100 \\
    Number of detectors in array & 16 & 216 \\
    Total active area [cm$^{2}$] & 116 & 216 \\
    Energy resolution [eV] & $\sim$150& $\sim$140 \\
    Time resolution FWHM [ns] & no & $\sim$400\\
    \hline\hline
  \end{tabular}
\end{table}

\begin{table}
  \centering
  \caption{The properties of kaonic hydrogen and kaonic deuterium atoms are shown.
  The values $\epsilon_{1s}$ and $\Gamma_{1s}$ of kaonic deuterium according to recent
  theoretical studies are given.}\label{KH-KD-P}
  \vspace{0.5 cm}
  \begin{tabular}{|l|l|l|}
    \hline\hline
    Kaonic atom & KH & KD \\
    \hline
    K$_{\alpha}$ e.m. value [eV] & 6480 & 7810 \\
    $\epsilon_{1s}$ [eV] & $\sim$ 200 & $\sim$ 325 \cite{Ivanov-KD} \\
    $\Gamma_{1s}$ [eV] & $\sim$ 250 & $\sim$ 630 \cite{Ivanov-KD}\\
    X-ray yield ( K$_{\alpha}$) [\%] & 1-3  & $\sim$ 0.2  \\
    \hline\hline
  \end{tabular}
\end{table}

An array of 12 subunits (total active area 216 cm$^{2}$) will
surround the cryogenic gas target. This target-detector system is
in development by the SIDDHARTA Collaboration
\cite{cargnelli05,marton-ect06,zmeskal-fb18} and will be setup at
DA$\Phi$NE in 2007.

\begin{figure}
  \includegraphics[height=.3\textheight]{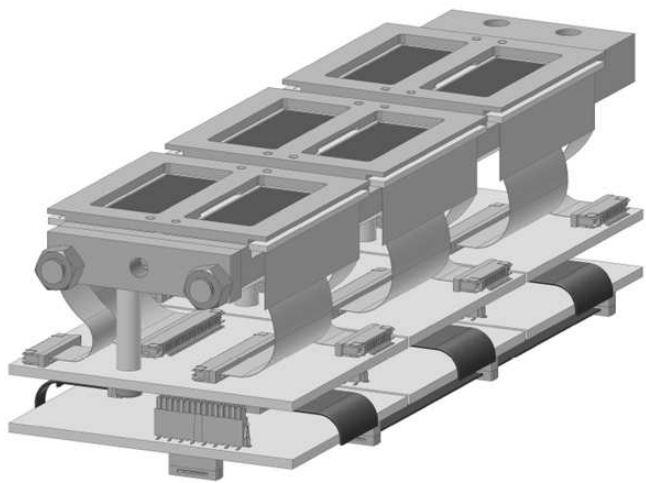}
  \caption{A SDD subunit assembled with 6 SDD chips (active area in total 18 cm$^{2}$).
  The setup of the X-ray detection system will consist of
  12 subunits.}
\end{figure}

\subsection{Setup development}

The SIDDHARTA setup was guided by the successful design of the
DEAR setup. The construction phase started in the second half of
2006 and will be finished in spring 2007. A first prototype of the
cryogenic target cell was developed and built. The light weight
cell is made only from selected materials (pure aluminium and
Kapton) and was successfully tested at a temperature of 25 K. The
material used for this setup, especially for the target cell, the
SDD mounting and the structure material in the vicinity of the SDD
chips were carefully analyzed by PIXE (Proton Induced X-ray
Emission) at VERA \cite{VERA}. The element content of the
materials in use was detected with a sensitivity for Fe impurities
in the ppm range. The iron content of the mechanical components
was always less than 50 ppm, as required by the experiment. Also
the ceramic frame of the SDD chip was analyzed and the ceramic
material with the lowest iron content was selected. The
installation of the whole experimental setup at DA$\Phi$NE is
planned for mid of 2007.

\begin{figure}
  \includegraphics[height=.3\textheight]{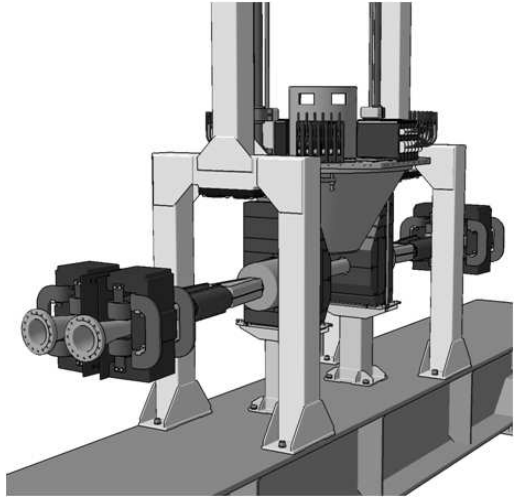}
  \caption{Schematic view of the SIDDHARTA setup at the intersection.}
\end{figure}

\subsection{Monte Carlo simulations for kaonic hydrogen and deuterium}

Monte Carlo simulations for the SIDDHARTA setup show that a
precision at the percent level can be anticipated for
$\epsilon_{1s}$ and $\Gamma_{1s}$ of kaonic hydrogen. More
challenging might be the measurement of the X-ray spectrum of
kaonic deuterium because of the experimentally unknown but as an
order of magnitude smaller anticipated K$_{\alpha}$ X-ray yield
(see tab.2). Furthermore, theoretical predictions on the strong
interaction width differ \cite{wycech}.

Assuming an average luminosity of 10$^{32}´$ cm$^{-2}$s$^{-1}$ in
a beam time of $\sim$60 days  >5$\cdot$10$^{4}$ kaonic hydrogen
K$_{\alpha}$ events can be collected - therefore sufficient to
fulfill the goal of a percent level measurement. In the kaonic
deuterium case a Monte Carlo simulation (see fig.3) shows that
measurement becomes feasible. In the SIDDHARTA experiment more
than 5000 kaonic deuterium K$_{\alpha}$ events can be collected in
60 days.

\begin{figure}
  \includegraphics[height=.3\textheight]{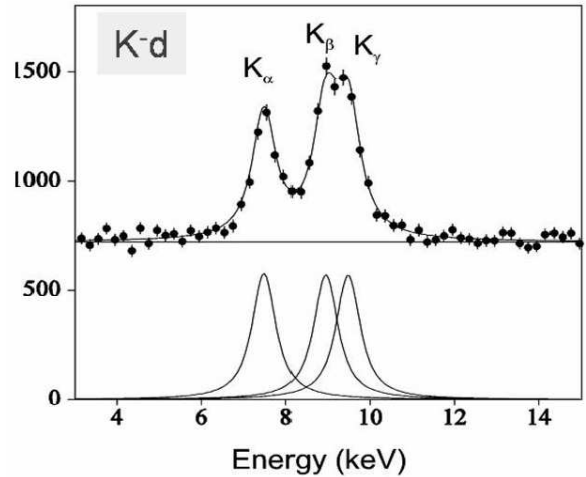}
  \caption{Monte Carlo simulation of the kaonic deuterium X-ray
  spectrum assuming the values: shift = -325 eV, width = 630 eV, X-ray yield = 0.2\%.
  In this case the shift and the width can be determined with a precision of <6\% and <8\%, respectively.}
\end{figure}

The Monte Carlo simulations show that even with an assumed width
of 1200 eV the values of $\epsilon_{1s}$ and $\Gamma_{1s}$ can be
measured with <10\% uncertainty.


\section{Outlook}

First test measurements with the new X-ray detection system and
the new setup are foreseen to take place in 2007.

At the highest priority measurements on kaonic hydrogen and kaonic
deuterium will be performed - these data are highly requested
\cite{catalina-cern}.

With the shift and width data of kaonic hydrogen and deuterium the
isospin-dependent scattering lengths can be extracted. New
experimental information about the kaon-proton interaction at
threshold and the elusive $\Lambda$(1405) resonance - important
for the research on deeply bound kaonic systems - will be
provided. Moreover, kaonic helium ($^{4}$He and $^{3}$He) will be
studied in the framework of SIDDHARTA.


\begin{theacknowledgments}

The work of the SIDDHARTA Collaboration is supported by EU
Integrated Infrastructure Initiative HadronPhysics with the the
Joint Research Activity JRA10 and TARI-INFN, contract number
RII3-CT-2004-506078.
\end{theacknowledgments}



\bibliographystyle{aipproc}   


\begin{thebibliography}{9}


\bibitem{Indelicato05} J.P.~Santos et al., Phys. Rev. \textbf{A71}, 032501 (2006).

\bibitem{PDG} J. Phys. G: Nucl. Part. Phys. \textbf{33}, 1027 (2006).

\bibitem{Weise95}N.~Kaiser, P.B.~Siegel, and W.~Weise, Nucl. Phys.
\textbf{A594}, 325 (1995).

\bibitem{akaishi} Y. Akaishi and T. Yamazaki, Phys. Rev. \textbf{C65},
044005 (2002).

\bibitem{Akaishi04}Y.Akaishi, Y.~Yamazaki and A.~Dot$\acute{e}$,
Nucl. Phys. \textbf{A738}, 168 (2004); Y.~Akaishi, Y. ~Yamazaki
and A.~Dot$\acute{e}$, Nucl. Phys. \textbf{A738}, 372 (2004).

\bibitem{FOPI} K. Suzuki, \emph{Proceedings of EXA05}, Eds. A. Hirtl, J.
Marton, E. Widmann, J. Zmeskal, Austrian Academy of Sciences
Press, Vienna (2005) pp. 83--90.

\bibitem{FINUDA} M. Agnello et al., Phys. Rev. Lett. \textbf{94}, 212303
(2005).

\bibitem{Amadeus-LOI} AMADEUS Collaboration, Study of deeply bound
kaonic nuclear states at DA$\Phi$NE2, Letter-of-Intent, 2006.

\bibitem{JPARC}H. Ohnishi for J-PARC E-15 experiment,
Proc. of the Workshop on Exotic hadronic atoms, deeply bound
kaonic nuclear states and antihydrogen, June 19-24, 2006, ECT*,
Trento, Italy, eds. C. Curceanu (Petrascu), A. Rusetsky, E.
Widmann, hep-ph/0610201, p46.

\bibitem{dearcase}S. Bianco et al.,
Rivista del Nuovo Cimento \textbf{22}, No.11, 1
(1999).

\bibitem{ishiwatari04} T.~Ishiwatari et al., Phys. Lett. \textbf{B593} (1-4), 48 (2004).

\bibitem{KpX} M.~Iwasaki et al., Phys. Rev. Lett. \textbf{78}, 3067 (1997); T.M. Ito et
al., Phys. Rev. C58, 2366 (1998).

\bibitem{Beer} G.~Beer et al., Phys. Rev. Lett. \textbf{94}, 212302 (2005).

\bibitem{theory1} B.~Borasoy,R.~Ni$\ss$ler and W.~Weise, Phys. Rev.
Lett. \textbf{94} 213401 (2005).

\bibitem{theory2} J.A.~Oller, J.~Prades and M. ~Verbeni, Phys. Rev. Lett. \textbf{95} 172502 (2005).

\bibitem{theory3} A.N.~Ivanov et al., Eur. Phys. J. \textbf{A21}
11 (2004).

\bibitem{theory4} U.-G.~Mei\ss ner, U.~Raha, and A. ~Rusetsky, Eur. Phys. J.
\textbf{C35} 349 (2004), hep-ph/0402261.

\bibitem{Ivanov-KD} A.N.~Ivanov et al., Eur. Phys. J. \textbf{A23} 79
(2005).

\bibitem{wycech} S.~Wycech, arXiv:nucl-th/0408066 v1, 2004.

\bibitem{cargnelli05}
M.~Cargnelli et al. \emph{Proceedings of EXA05}, Eds. A. Hirtl, J.
Marton, E. Widmann, J. Zmeskal, Austrian Academy of Sciences
Press, Vienna (2005) pp. 313--320.

\bibitem{marton-ect06} J.~Marton for the DEAR/SIDDHARTA Collaborations,
Proc. of the Workshop on Exotic hadronic atoms, deeply bound
kaonic nuclear states and antihydrogen, June 19-24, 2006, ECT*,
Trento, Italy, eds. C. Curceanu (Petrascu), A. Rusetsky, E.
Widmann, hep-ph/0610201, p17.

\bibitem{zmeskal-fb18} J.~Zmeskal et al., Proc. of the 18th International
IUPAP Conference on Few Body Problems in Physics FB18, August 21 -
26, 2006, Santos (Sao Paulo), Brasil, submitted to Nucl. Phys.

\bibitem{VERA} Vienna Environmental Research Accelerator (VERA), University of
Vienna, Institut fur Isotopenforschung und
Kernphysik,http://www.univie.ac.at/Kernphysik/VERA/.

\bibitem{catalina-cern} C.~Curceanu, A.~Rusertski, E.~Widmann, Exotic
atoms cast light on fundamental questions, CERN Courier
\textbf{46}, 2006.

\end{thebibliography}




\end{document}